\documentstyle[aps,pre,preprint,epsf]{revtex}
\begin{document}
\def\OP {\tensor P}
\def\B.#1{{\bbox{#1}}}
\renewcommand{\thesection}{\arabic{section}}
\title{Stability Analysis of Flame Fronts: Dynamical Systems Approach in
the Complex Plane}
\author {Oleg Kupervasser, Zeev Olami and Itamar Procaccia}
\address{Department of~~Chemical Physics, The Weizmann
Institute of Science, Rehovot 76100, Israel} \maketitle
\begin{abstract}
We consider flame front propagation in channel geometries. The steady state
solution in this problem is space dependent, and therefore the linear
stability analysis is
described by a partial integro-differential equation with a space dependent
coefficient. Accordingly it involves complicated eigenfunctions. We show
that the analysis
can be performed to required detail using a finite order dynamical
system in
terms of the dynamics of singularities in the complex plane, yielding
detailed understanding
of the physics of the eigenfunctions and eigenvalues.
\end{abstract}
\section{Introduction}
In this paper we discuss the stability of steady flame fronts in channel
geometry. Traditionally \cite{Pel,BS,Vic}
one studies stability by considering the linear operator which is obtained
by linearizing the
equations of motion around the steady solution. The eigenfunctions obtained
are {\em delocalized}
and in certain cases are not easy to interpret.
In the case of flame fronts the steady state solution is space dependent
and therefore the
eigenfunctions are very different from simple Fourier modes. We show in
this paper that a good
understanding of the
nature of the eigenspectrum and eigenmodes can be obtained by doing almost
the opposite
of traditional stability analysis, i.e., studying the {\em localized}
dynamics of
singularities in the complex plane. By reducing the stability analysis to a
study of a finite
dimensional dynamical system one can gain considerable intuitive
understanding of the nature
of the stability problem.

The analysis is based on the understanding that for a given channel width
$L$ the steady state solution for the
flame front is given in terms of $N(L)$ poles that are organized on a line
parallel to the
imaginary axis \cite{85TFH}. Stability of this solution can then be
considered in two
steps. In the
first step we examine the response of this set of $N(L)$ poles to
perturbations in their
positions. This procedure yields an important part of the stability
spectrum. In the second
step we examine general perturbations, which can also be described by the
addition of extra poles to the
system of $N(L)$ poles. The response to these perturbations gives us the
rest of the stability
spectrum; the combinations of these two steps rationalizes all the
qualitative features found by traditional stability analysis.

In Sec.2 we present a brief review of the stationary solutions of front
propagation in
channel geometries. In Sec.3 we present the results of traditional linear
stability analysis,
and show the eigenvalues and eigenfunctions that we want to interpret by
using the pole
decomposition. Sec. 4 presents the analysis in terms of complex
singularities, in two steps as discussed above. A summary and discussion
is presented in Sec.5.

\section{Flame propagation in channel geometry}

We consider a channel of transverse width $2\pi L$, and of infinite extent
in the longitudinal direction.
A graph of a flame front propagating in this channel is defined as
$h(\theta,t)$ where $\theta$
is a rescaled transversal coordinate $0<\theta<2\pi$. The equation of motion
is given
conveniently in terms of $u(\theta,t)\equiv \partial h(\theta,t)/\partial
\theta$\cite{77Siv,90GS,96KOP,97OGKP}:
\begin{eqnarray}
&&{\partial u(\theta,t) \over \partial t}= {u(\theta,t)\over L^2}{\partial
u(\theta,t) \over \partial \theta }
+{\nu\over L^2}{\partial^2 u(\theta,t)\over \partial \theta^2}\nonumber\\
&&+ {1\over L}I\{u(\theta,t)\}, \label{eqfinal}
\end{eqnarray}
Here $\nu$ is a viscosity-like parameter and the functional
$I[u(\theta,t)]$ is conveniently
defined in terms of the spatial Fourier transform
\begin{eqnarray}
&&u(\theta,t)= \int_{-\infty}^{\infty} e^{i k x}\hat u(k,t) dk \ ,
\label{Four}\\
&& I[u(k,t)] = |k| \hat u(k,t) \ . \label{hil} \end{eqnarray}
It is very useful \cite{85TFH,82LC,89Jou,90Jou} to discuss the solutions of
these equations of
motion in terms of expansions in
$N$ poles whose position $z_j(t)\equiv x_j(t)+iy_j(t)$ in the complex plane
is time dependent:
\begin{eqnarray}
&&u(\theta,t)=\nu\sum_{j=1}^{N}\cot \left[{\theta-z_j(t) \over 2}\right]
+ c.c.\nonumber \\
&&=\nu\sum_{j=1}^{N}{2\sin [\theta-x_j(t)]\over \cosh [y_j(t)]-\cos
[\theta-x_j(t)]}\ , \label{upoles}
\end{eqnarray}

Substituting (\ref{upoles}) in (\ref{eqfinal}) we derive the following
ordinary differential
equations for the positions of the poles:
\begin{eqnarray}
&&-L^2{dx_{j}\over dt}=\nu\sum_{k=1,k\neq j}^N
\sin(x_j-x_k)\Bigg[
\Big[\cosh (y_j-y_k) \label{xj} \\ &&-\cos (x_j-x_k)\Big]^{-1}+\Big[\cosh
(y_j+y_k)-\cos (x_j-x_k)
\Big]^{-1}\Bigg] \nonumber \\
&& L^2{dy_{j}\over dt}=\nu\sum_{k=1,k\neq j}^{N }\Big({\sinh (y_j-y_k)\over
\cosh (y_j-y_k) -\cos (x_j-x_k)} \nonumber \\ &&+ {\sinh (y_j+y_k)\over
\cosh (y_j+y_k)
-\cos (x_j-x_k)} \Big)+\nu\coth (y_j)- L \label{yj} .
\end{eqnarray}
In particular we can find the steady state solution $u_s(\theta)$ by demanding
$\dot x_j=\dot y_j=0$ and stability. The solution is
\begin{equation}
u_s(\theta)=\nu\sum_{j=1}^{N}{2\sin [\theta-x_s]\over \cosh [y_j]-\cos
[\theta-x_s]}\ , \label{stat}
\end{equation}
where $x_s$ is the real (common) position of the stationary poles and $y_j$
their
stationary imaginary position. We need
to determine the actual positions $y_j$. This is done numerically by
running the equations of motion
for the poles starting from $N$ poles in initial positions and
waiting for relaxation.
A complete analysis of this steady-state solution was first presented in
Ref.\cite{85TFH} and the main
results are summarized as follows:
\begin{enumerate}
\item There is only one
stable stationary solution which is geometrically represented by a giant
cusp (or equivalently one finger) and
analytically by $N(L)$ poles which are aligned on one line parallel to the
imaginary
axis. The existence of this solution is made clearer with the following
remarks.
\item There exists an attraction between the poles along the real line. This is
obvious from Eq.(\ref{xj}) in which the sign of $dx_j/dt$ is
always determined by $\sin(x_j-x_k)$. The resulting dynamics merges all the $x$
positions of poles whose $y$-position remains finite. \item The $y$
positions are distinct, and the poles are aligned above each others in
positions
$y_{j-1}<y_j<y_{j+1}$ with the maximal being $y_{N(L)}$. This can be
understood from
Eq.(\ref{yj}) in which the interaction is seen to be repulsive at short ranges,
but changes sign at longer ranges.
\item If one adds
an additional pole to such
a solution, this pole (or another) will be pushed to infinity along the
imaginary axis.
If the system has less than $N(L)$ poles it is unstable to the addition of
poles,
and any noise will drive the system towards this unique state. The number
$N(L)$ is
\begin{equation}
N(L)= \Big[{1 \over 2}\left( {L \over \nu }+1\right) \Big]\ , \label{NofL}
\end{equation}
where $\Big[ \dots \Big]$ is the integer part. To see this consider a
system with $N$ poles and such
that all the values of $y_j$ satisfy the condition $0< y_j <y_{max}$. Add
now one
additional pole whose coordinates are $z_a\equiv (x_a,y_a)$ with $y_a\gg
y_{max}$. From the equation of motion for $y_a$, (\ref{yj}) we see that the
terms in
the sum are all of the order of unity as is also the $\cot(y_a)$ term. Thus
the equation of motion of $y_a$ is approximately
\begin{equation}
{dy_a \over dt}\approx \nu{2N+1 \over L^2}-{1\over L} \ . \label{ya}
\end{equation}
The fate of this pole depends on the number of other poles. If $N$ is too
large the pole
will run to infinity, whereas if $N$ is small the pole
will be attracted towards the real axis. The condition for moving away to
infinity
is that $N > N(L)$ where $N(L)$ is given by (\ref{NofL}). On the other hand
the $y$
coordinate of the poles cannot hit zero. Zero is a repulsive line, and poles
are pushed away from zero with infinite velocity.
To see this consider
a pole whose $y_j$ approaches zero. For any finite $L$ the term
$\coth(y_j)$ grows
unboundedly whereas all the other terms in Eq.(\ref{yj}) remain bounded.
\item The height of the cusp is proportional to $L$. The distribution of
positions
of the poles along the line of constant $x$ was worked out in \cite{85TFH}.
\end{enumerate}
We will refer to the solution with all these properties as the
Thual-Frisch-Henon
(TFH)-cusp solution.

\section{Linear Stability Analysis in Channel Geometry}

The standard technique to study the linear stability of the steady solution
is to
perturb it by a
small perturbation $\phi(\theta,t)$: $u(\theta,t) =
u_s(\theta)+\phi(\theta,t)$ .
Linearizing the dynamics for small $\phi$ results in the following equation
of motion
\begin{eqnarray}
{\partial \phi(\theta,t) \over \partial t}&=& {1\over L^2}\Big
[\partial_\theta [u_s(\theta) \phi(\theta,t)]
\nonumber \\&+&\nu \partial_\theta^2\phi(\theta,t)\Big]
+{1\over L}I(\phi(\theta,t)) \ . \label{linear} \end {eqnarray}
were the linear operator contains $u_s(\theta)$ as a coefficient. Accordingly
simple Fourier
modes do not diagonalize it. Nevertheless, we proceed to decompose
$\phi(x)$ in Fourier
modes according to ,
\begin{eqnarray}
\phi(\theta,t)&=&\sum_{k=-\infty}^{\infty} \hat\phi_k(t)
e^{ik\theta}\label{phi}\\
u_s(\theta)&=&-2{\nu}i\sum_{k=-\infty}^{\infty}
\sum_{j=1}^N sign(k)e^{-\mid k \mid y_j}e^{ik\theta}
\end {eqnarray}
The last equation follows from (\ref{stat}) by expanding in a series of
$\sin{k\theta}$. In these sums the discrete $k$ values run over all the
integers. Substituting in Eq.(\ref{linear}) we get:
\begin{equation}
{d\hat \phi_k(t) ) \over dt}= \sum _n a_{kn} \hat\phi_n(t)\ , \
\end {equation}
where $a_{kn}$ are entires of an infinite matrix:
\begin{eqnarray}
a_{kk}&=&{\mid k \mid\over L} -{\nu \over L^2} k^2 \ , \label{akk}\\ a_{kn}
&=&{k\over L^2}sign(k-n)({2\nu} \sum_{j=1}^N e^{-\mid k-n \mid y_j})
\quad k \neq n \ .
\end {eqnarray}
To solve for the eigenvalues of this matrix we need to truncate it at some
cutoff
$k$-vector
$k^*$. The scale $k^*$ can be chosen on the basis of Eq.(\ref{akk}) from
which we see
that the largest value of $k$ for which $a_{kk}\ge 0$ is a scale that we denote
as $k_{max}$, which is the integer part of $L/\nu$.
We must choose $k^*>k_{max}$ and test the choice by the convergence of the
eigenvalues.
The chosen value of $k^*$ in our numerics was $4k_{max}$. One should notice
that this cutoff limits the number of eigenvalues, which should be
infinite. However
the lower eigenvalues will be well represented. The results for the low order
eigenvalues of the matrix $a_{kn}$ that were obtained
from the converged numerical calculation are presented in Fig.1

The eigenvalues are
multiplied by $L^2/\nu$ and are plotted as a function of $L$. We order the
eigenvalues in
decreasing order and denote them as $\lambda_0\ge \lambda_1\ge \lambda_2 \dots$.
The figure offers a number of qualitative observations: \begin{enumerate}
\item There exists an obvious Goldstone or translational mode
$u'_s(\theta)$ with eigenvalue $\lambda_0=0$.
This eigenmode stems from the Galilean invariance of the equation of motion.
\item The eigenvalues oscillate periodically between values that are
$L$-independent in this
presentation (in which we multiply by $L^2$). In other words, up to the
oscillatory
behaviour the eigenvalues depend on $L$ like $L^{-2}$.
\item The eigenvalues
$\lambda_1$ and $\lambda_2$ hit zero periodically. The functional
dependence in this presentation appears almost piece-wise linear.
\item The higher eigenvalues are more negative. They exhibit similar
qualitative
behaviour, but without reaching zero.
We note that the solution becomes marginally stable for every value of $L$
for which the eigenvalues
$\lambda_1$ and $\lambda_2$ hit zero. The $L^{-2}$ dependence of the
spectrum indicates
that the solution becomes more and more sensitive to noise as $L$ increases
\cite{98kop}.
\end{enumerate}
In addition to the eigenvalues, the truncated matrix also yields
eigenvectors that we
denote as $\B.A^{(\ell)}$. Each such vector has $k^*$ entries, and we can
compute the eigenfunctions $f^{(\ell)}(\theta)$ of the linear operator
(\ref{linear}),
using (\ref{phi}), as
\begin{equation}
f^{(\ell)}(\theta) \equiv \sum _{-k^*}^{k^*} e^{ik\theta} A^{(\ell)}_k \ .
\label{fell}
\end{equation}
Eq.(\ref {linear}) does not mix even with odd solutions in $\theta$, as can
be checked
by inspection. Consequently the available solutions have even or odd parity,
expandable in either $\cos$ or $\sin$ functions. The first two
nontrivial eigenfunctions $f^{(1)}(\theta)$ and $f^{(2)}(\theta)$ are
shown in Figs.2,3. It is evident that the function in Fig.2 is odd around
zero whereas in Fig.3 it is even. Similarly we can numerically generate any
other eigenfunction of the linear operator,
but we understand neither the physical significance of these eigenfunction
nor the $L$ dependence of their associated eigenvalues shown in Fig.1
In the next section we
will demonstrate how the dynamical system approach in terms of
singularities in the
complex plane provides us with considerable intuition about these issues.
\section{Linear Stability in terms of complex singularities} Since the
partial differential equation
is continuous there is an infinite number of modes. To understand this in
terms of pole dynamics we consider
the problem in two steps: First, we consider the $2N(L)$ modes associated
with the dynamics of the
$N(L)$ poles of the giant cusp.
In the second step we explain
that all the additional modes result from the introduction of additional
poles, including
the reaction of the $N(L)$ poles of the giant cusp to the new poles. After
these two steps
we will be able to identify all the linear modes that were found
by diagonalizing the stability matrix in the previous section.
\subsection{The modes associated with the giant cusp} In the steady
solution all the
poles occupy stable equibirilium positions. The forces operating
on any given pole cancel exactly, and we can write matrix equations for
small perturbations
in the pole positions $\delta y_i$ and $\delta x_i$.

Following \cite{85TFH} we rewrite the equations of motion (\ref{yj}) using
the Lyapunov function $U$:
\begin{equation}
L\dot{y_i}={\partial U \over \partial y_i} \label{hil1} \end{equation}
where ${i=1,...,N}$ and

\begin{eqnarray}
U={\nu \over L}[ \sum_i\ln \sinh y_i&+&2\sum_{i<k}
(\ln \sinh{y_k-y_i \over 2}\nonumber \\&+&\ln \sinh{y_k+y_i \over 2})] -
\sum_iy_i \label{hil2}
\end{eqnarray}
The linearized equations of motion for $\delta y_i$ are: \begin{equation}
L\dot{\delta y_i}= \sum_k{\partial^2 U \over \partial y_i \partial y_k
}\delta y_k \ . \label{linyj}
\end{equation}
The matrix $\partial^2 U/\partial y_i \partial y_k$ is real and symmetric
of rank $N$.
We thus expect to find $N$ real eigenvalues and N orthogonal eigenvectors.

For the deviations $\delta x_i$ in the $x$ positions we find the following
linearized equations of
motion
\begin{eqnarray}
&&L\dot{\delta x_j}=-{\nu \over L}{\delta x_j} \sum_{k=1, k \neq j}^N
({1 \over \cosh(y_j-y_k)-1)}\nonumber \\ && +{1 \over \cosh(y_j+y_k)-
1}) \nonumber \\ &&+
{\nu \over L} \sum_{k=1, k \neq j}^N {\delta x_k}
({1 \over \cosh(y_j-y_k)-1)}+{1 \over \cosh(y_j+y_k)- 1})
\end{eqnarray}
In shorthand:
\begin{equation}
L{d{\delta x_i}\over dt}=V_{ik}\delta x_k \ . \label{linxi} \end{equation}
The matrix $V$ is also real and symmetric. Thus $V$ and ${\partial^2 U /
\partial y_i \partial y_k}$
together supply $2 N(L)$ real eigenvalues and $2N(L)$ orthogonal
eigenvectors. The explicit form of
the matrices $V$ and ${\partial^2 U / \partial y_i \partial y_k}$ is as follows:
For $i\neq k$:
\begin{equation}
{\partial^2 U \over \partial y_i \partial y_k}= {\nu \over L}[{1/2 \over
\sinh^2({y_k-y_i \over 2})}-{1/2 \over\sinh^2({y_k+y_i \over 2})}]
\label{mat11}
\end{equation}

\begin{equation}
V_{ik}={\nu \over L}({1 \over \cosh(y_i-y_k)-1)}+{1 \over \cosh(y_i+y_k)- 1})
\label{mat12}
\end{equation}

and for $i=k$ one gets:
\begin{eqnarray}
{\partial^2 U \over \partial y_i^2}=
&-&{\nu \over L}[\sum_{k \neq i}^N
\left({1 \over 2\sinh^2({y_k-y_i \over 2})} +{1 \over 2\sinh^2({y_k+y_i
\over 2})}\right)
\nonumber \\&+&{1 \over \sinh^2(y_i)}]
\label{mat21}
\end{eqnarray}
\begin{equation}
V_{ii}=\sum_{k \neq i}^N[-{\nu \over L}
({1 \over \cosh(y_i-y_k)-1)}+{1 \over \cosh(y_i+y_k)-
1})]
\label{mat22}
\end{equation}
Using the known steady state solutions $y_i$ at any given $L$ we can
diagonalize
the $N(L)\times N(L)$ matrices numerically. In  Fig.4 we present
the eigenvalues of the lowest order
modes obtained from this procedure. The least negative eigenvalues
touch zero periodically. This eigenvalue can be fully identified with the
motion of the highest
pole $y_{N(L)}$ in the giant cusp. At isolated values of $L$ the
position of this pole tends
to infinity, and then the row and the column in our matrices that contain
$y_{N(L)}$ vanish identically, leading to a zero eigenvalue. The rest of the
upper eigenvalues match perfectly with half of the observed eigenvalues in
Fig.1. In other words, the eigenvalues observed here agree perfectly
with the ones plotted in this Fig.1 until the discontinuous increase
from their minimal points.
The ``second half'' of the oscillation in the eigenvalues as a function of
$L$ is not contained in this spectrum of the $N(L)$ poles of the giant cusp.
To understand the rest
of the spectrum we need to consider perturbation of the giant cusp by
additional poles.
The eigenfunctions can be found using the knowledge of the eigenvectors of
these matrices.
Let us denote the eigenvectors of ${\partial^2 U / \partial y_i \partial
y_k}$ and $V$ as
$\B.a^{(\ell)}$ and $\B.b^{(\ell)}$ respectively. The perturbed solution is
explicitly given as (taken for $x_s=0$): \begin{equation}
u_s(\theta)+\delta u=2\nu \sum_{i=1}^N {\sin(\theta -\delta x_i) \over
\cosh(y_i+\delta y_i)- \cos (\theta -\delta x_i)} \end{equation}
where $\delta u$ is
\begin{eqnarray}
\delta u=&-&4{\nu }\sum_{i=1}^N \sum_{k=1}^{\infty} \delta y_i
ke^{-ky_i}\sin k\theta
\nonumber\\ &-&4\nu \sum_{i=1}^N \sum_{k=1}^{\infty}
\delta x_i ke^{-ky_i}\cos k\theta \label{hil4}
\end{eqnarray}
So knowing the eigenvectors $\B.a^{(\ell)}$ and $\B.b^{(\ell)}$ we can
estimate the
eigenvectors $f^{(\ell)}(\theta)$ of (\ref{fell}): \begin{equation}
f^{(\ell)}_{\rm sin}(\theta) =-4{\nu }\sum_{i=1}^N \sum_{k=1}^{\infty}
a_i^{(j)} k e^{-k y_i}\sin k\theta \ , \quad j=1,...,N
\label{hhh1}
\end{equation}
or
\begin{equation}
f^{(\ell)}_{\rm cos}(\theta) =-4{\nu }\sum_{i=1}^N \sum_{k=1}^{\infty}
b_i^{(j)} ke^{-ky_i}\cos k\theta\ , \quad j=1,...,N
\label{hhh2}
\end{equation}
where we display separately the $\sin$ expansion and the $\cos$ expansion.
For the case $j=1$,
the eigenvalue is zero, and a uniform translation of the poles in any
amount $\delta x_i$ results
in a Goldstone mode. This is characterized by an eigenvector $b^{(1)}_i=1$
for all $i$.
The eigenvectors $f^{(\ell)}$ (Fig.5,6)computed this way are identical to
numerical
precision with those shown in Figs.2,3, and observe the agreement.

\subsection{Modes related to additional poles} In this subsection we
identify the
rest of the modes that were not found in the previous
subsection. To this aim we study the response of the TFH solution
to the introduction
of additional poles. We choose to add $M$ new poles all positioned at the
same imaginary
coordinate $y_p\ll y_{max}$, distributed at equidistant real positions
$\{x_j=x_0+(2\pi/M)j\}_{j=1}^M$.
For $x_0=0$ we use (\ref{upoles}) and the Fourier expansion  to
obtain a perturbation of the form
\begin{equation}
\delta u(\theta,t) \simeq 4{\nu }M e^{-M y_p (t)}\sin M\theta \label{hil5}
\end{equation}
For $x_0=-\pi/2M$ we get
\begin{equation}
\delta u(\theta,t) \simeq 4{\nu }Me^{-My_p (t)}\cos M\theta \label{hil6}
\end{equation}
in both cases the equations for the dynamics of $y_p$ follow from
Eqs.(\ref{xj})-(\ref{yj}):
\begin{equation}
{dy_p \over dt} \simeq 2{\nu \over L^2}\alpha(M) \ , \label{dyp} \end{equation}
where $\alpha(M)$ is given as:
\begin{equation}
\alpha(M)=[{1 \over 2}({L \over \nu}+1)]-{1 \over 2}({L \over \nu}-M)
\end{equation}
Since (\ref{dyp}) is linear, we can solve it and substitute in
Eqs.(\ref{hil5})-(\ref{hil6}).
Seeking a form $\delta u(\theta,t)\sim exp(-\lambda(M) t)$ we find that the
eigenvalue $\lambda(M)$ is
\begin{equation}
\lambda(M) =2M{\nu \over L^2}\alpha(M)
\label{lamm}
\end{equation}
These eigenvalues are plotted in Fig.7
At this point we consider the dynamics of the poles in the giant cusp under
the influence
of the additional $M$ poles. From
Eqs.(\ref{linyj}), (\ref{linxi}), (\ref{xj}), (\ref{yj}) we obtain, after some
obvious algebra,
\begin{equation}
L\dot{\delta y_i}= \sum_j{\partial^2 U \over \partial y_i \partial y_j
}\delta y_j-4{\nu \over L}Me^{-My_p(t)} \sinh(My_i) \end{equation}
or
\begin{equation}
L\dot{\delta x_i}= \sum_j{V_{ij}
}\delta x_j-4{\nu \over L}Me^{-My_p(t)} \cosh(My_i) \end{equation}
It is convenient now to transform from the basis $\delta y_i$ to the
natural basis
$w_i$ which is obtained using the linear transformation $\B.w =A^{-1}
\B.{\delta y}$.
Here the matrix $A$ has columns which are the eigenvectors of $\partial^2 U
/\partial y_i \partial y_j$
which were computed before. Since the matrix was real symmetric, the matrix
$A$ is
orthogonal, and $A^{-1}=A^T$. Define $C=4{\nu \over L^2}M e^{-M y_p(0)}$
and write
\begin{equation}
\dot{ w_i}=-\lambda_i w_i-Ce^{-\lambda(M) t} \xi_i \ , \label{wi} \end{equation}
where $-\lambda_i$ are the eigenvalues associated with the columns of $A$, and
\begin{equation}
\xi_i=\sum_j A_{ji}\sinh M y_j \ .
\end{equation}
We are looking now for a solution that decays exponentially at the rate
$\lambda(M)$:
\begin{equation}
w_i(t)=w_{i}(0)e^{-\lambda(M) t}
\end{equation}
Substituting the desired solution in (\ref{wi}) we find a condition on the
initial value of $w_i$:
\begin{equation}
w_{i}(0)=-{C \over \lambda_i- \lambda(M)}\xi_i \end{equation}
Transforming back to $\delta y_i$ we get
\begin{eqnarray}
&&	\delta y_{i}(0)=\sum_k A_{ik}w_{k}(0)=-\sum_k A_{ik}
{C \over \lambda_k- \lambda(M)}\sum_l A_{lk} \sinh M y_l \nonumber \\&&
=-C\sum_l\sinh M y_l \sum_k{A_{ik} A_{lk} \over \lambda_k- \lambda_p^M}
\label{hil7}
\end{eqnarray}

We can get the eigenfunctions of the linear operator, as before, using
Eqs.(\ref{hil4}),
(\ref{hil5}), (\ref{hil6}), (\ref{hil7}). We get \begin{eqnarray}
&&f^{(M)}_{\rm sin}(\theta) =4C {\nu }\sum_{i=1}^{N(L)} \sum_{k=1}^{\infty}
(\sum_l\sinh M y_l \sum_m{A_{im} A_{lm} \over \lambda_m- \lambda(M)})
\nonumber \\&&
\times ke^{-ky_i}\sin k\theta+{L^2 }C\sin M\theta \
\label{ril7}
\end{eqnarray}
An identical calculation to the one started with Eq. (\ref{wi}) can be
followed for
the deviations $\delta x_i$. The final result reads \begin{eqnarray}
&& f^{(M)}_{\rm cos}(\theta) =4C {\nu }\sum_{i=1}^{N(L)} \sum_{k=1}^{\infty}
(\sum_l\cosh My_l \sum_m{\tilde A_{im} \tilde A_{lm} \over \tilde\lambda_m-
\lambda(M)}) \nonumber \\&&
\times ke^{-ky_i}\cos k\theta
+{L^2 }C\cos M\theta \ ,
\label{ril8}
\end{eqnarray}
where $\tilde A$ is the matrix whose columns are the eigenvectors of $V$,
and $-\tilde \lambda_i$
its eigenvalues.

We are now in position to explain the entire linear spectrum using the
knowledge that we have gained.
The spectrum consists of two separate types of contributions. The first type
has $2N$ modes that belong to the dynamics of the unperturbed $N(L)$ poles
in the giant cusp.
The second part, which is most of the spectrum, is built from modes of the
second type since $M$ can go to infinity. This structure is seen in the
Fig.4 and Fig.7.

We can argue that the set of eigenfunctions obtained above is complete and
exhaustive. To do this we show that any arbitrary periodic function of $\theta$
can be expanded in terms of these eigenfunctions. Start with the standard
Fourier series in terms of sin and cos functions. At this point solve
for $\sin{k\theta}$ and $\cos{k\theta}$ from Eqs.(\ref{ril7}-\ref{ril8}).
Substitute the results in the Fourier sums. We now have an expansion in terms
of the eigenmodes $f^{(M)}$ and in terms of the triple sums. The triple sums
however can be expanded, using Eqs.(\ref{hhh1}-\ref{hhh2}), in terms of the
eigenfunctions $f^{(\ell)}$. We can thus decompose any function in terms of the
eigenfunctions $f^{(M)}$ and $f^{(\ell)}$.
\section{Conclusions}
We discussed the stability of flame fronts in channel geometry using the
representation of the
solutions in terms of singularities in the complex plane. In this language
the stationary solution, which is a giant cusp in configuration space, is
represented by $N(L)$ poles which are organized on a line parallel to the
imaginary axis. We showed that the stability problem can be understood in
terms of two types of perturbations. The first type is a perturbation in the
positions of the poles that make up the giant cusp. The longitudinal motions
of the poles give rise to odd modes, whereas the transverse motions to even
modes. The eigenvalues associated with these
modes are eigenvalues of a finite, real and symmetric matrices, cf.
Eqs.(\ref{mat11}),
(\ref{mat12}), (\ref{mat21}), (\ref{mat22}). The second type of perturbations
is obtained
by adding  poles to the set of $N(L)$ poles representing the giant cusp.
The reaction of the latter poles is again separated into odd and even functions
as can be seen from Eqs.(\ref{hil5}),
(\ref{hil6}). Together the two types of perturbations rationalize and
explain all the features of the eigenvalues and eigenfunctions obtained from
the standard linear stability analysis.

\acknowledgments
This work has been supported in part by the Israel Science Foundation
administered
by the Israel Academy of Sciences and Humanities.

\begin{figure}
\epsfxsize=9.0truecm
\epsfbox{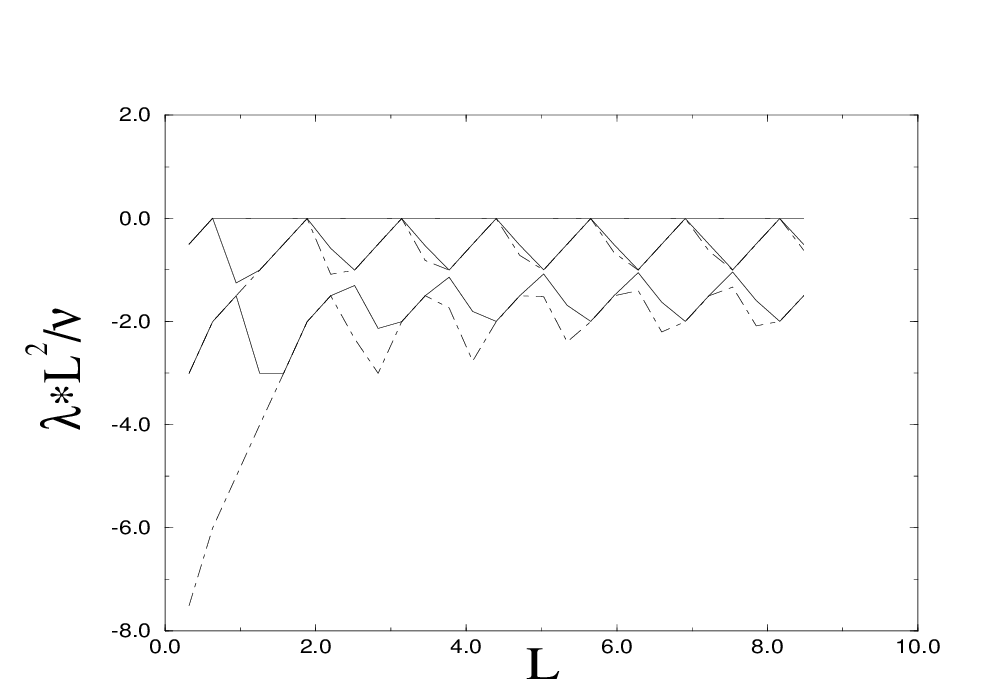}
\caption
{A plot of the first five eigenvalues obtained by diagonalizing the
matrix obtained by
traditional stability analysis, against the system size.
The eigenvalues are normalized by $L^2/\nu$. The largest eigenvalue is zero,
which is a Goldstone mode. All the other eigenvalues are negative
except for the second and third that touch zero periodically.
The second and fourth eigenvalues are represented by a  solid line and the third
and fifth eigenvalues are represented by a dot-dashed line.}
\label{file=Fig.1}
\end{figure}
\begin{figure}
\epsfxsize=9.0truecm
\epsfbox{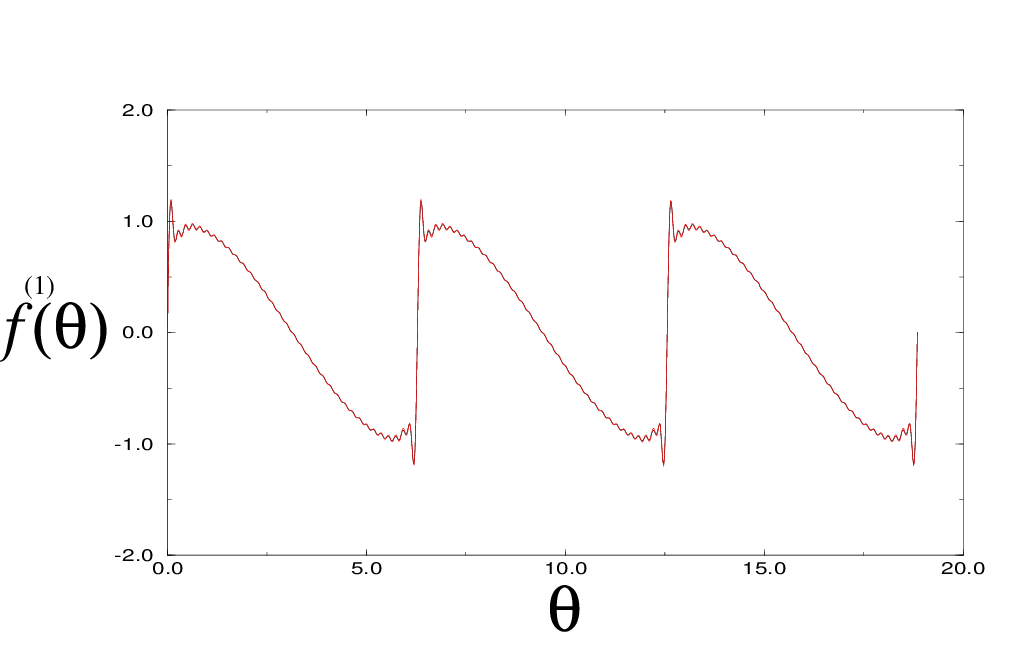}
\caption
{The first odd eigenfunction obtained from traditional stability analysis.}
\label{file=Fig.2}
\end{figure}
\begin{figure}
\epsfxsize=9.0truecm
\epsfbox{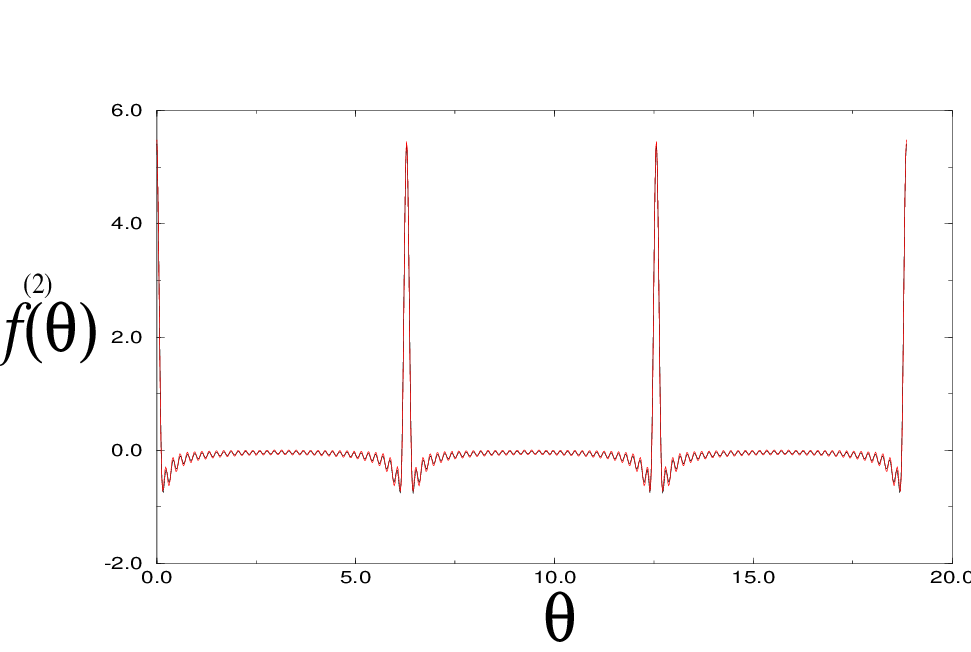}
\caption
{The first even eigenfunction obtained from traditional stability analysis.}
\label{file=Fig.3}
\end{figure}
\begin{figure}
\epsfxsize=9.0truecm
\epsfbox{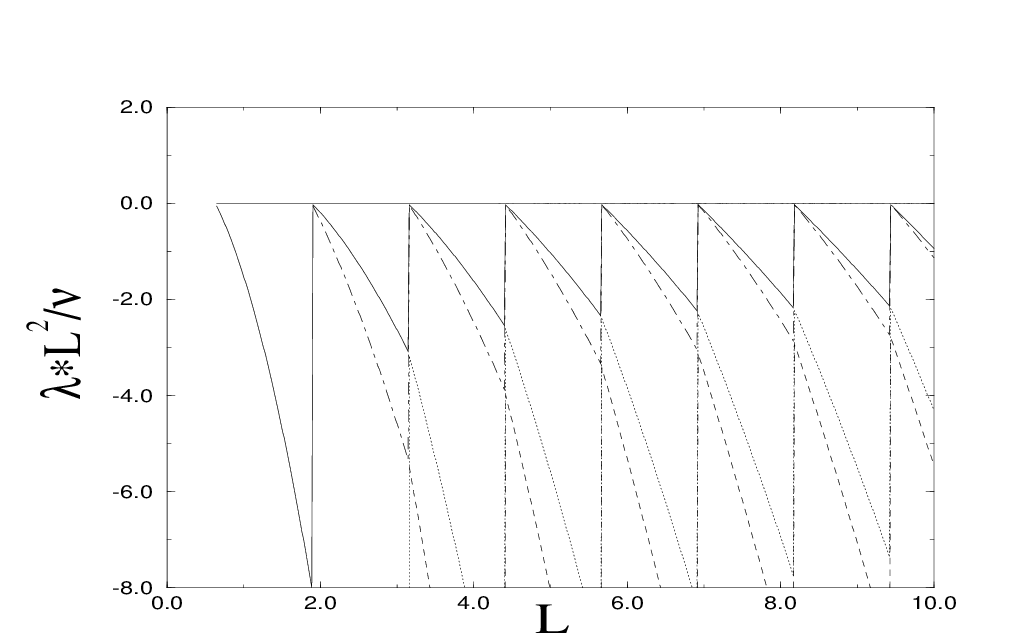}
\caption
{The eigenvalues associated with perturbing the positions of the
poles that consist the giant cusp.
The largest eigenvalue is zero. The second, third, fourth and fifth eigenvalues
are represented by
a solid line, dot-dashed line, dotted line and dashed line respectively.}
\label{file=Fig.4}
\end{figure}
\begin{figure}
\epsfxsize=9.0truecm
\epsfbox{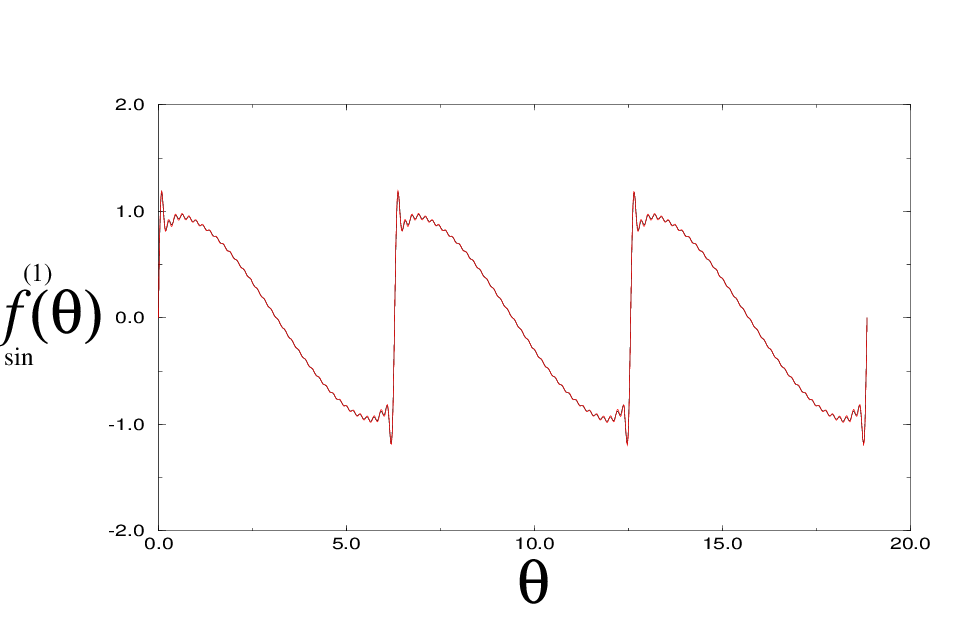}
\caption
{The first odd eigenfunction associated with perturbing the positions
of the poles in the giant cusp.}
\label{file=Fig.5}
\end{figure}

\begin{figure}
\epsfxsize=9.0truecm
\epsfbox{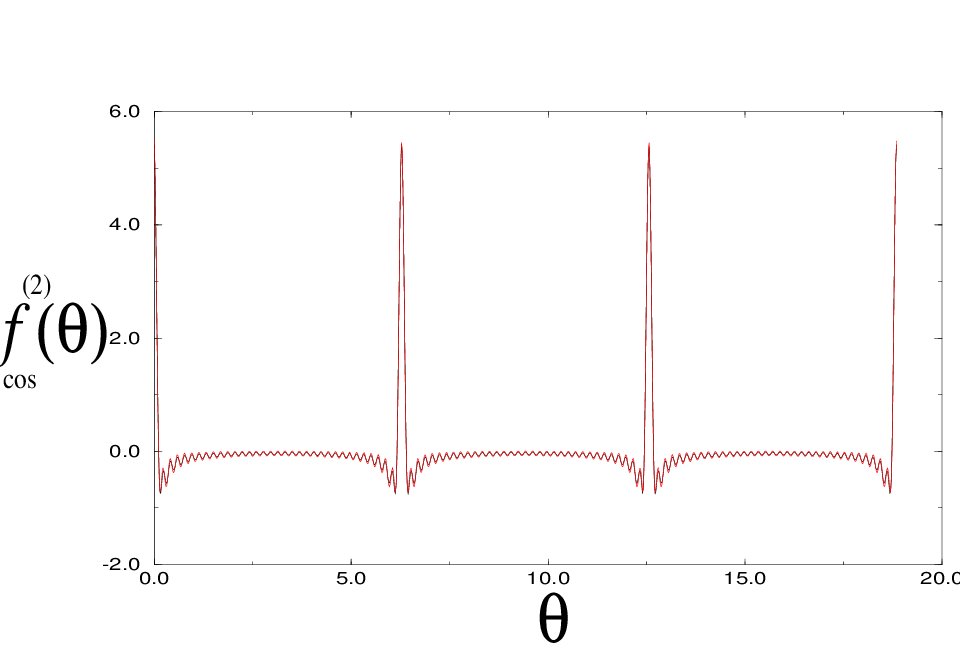}
\caption
{The first even eigenfunction associated with perturbing the positions
of the poles in the giant cusp.}
\label{file=Fig.6}
\end{figure}
\begin{figure}
\epsfxsize=9.0truecm
\epsfbox{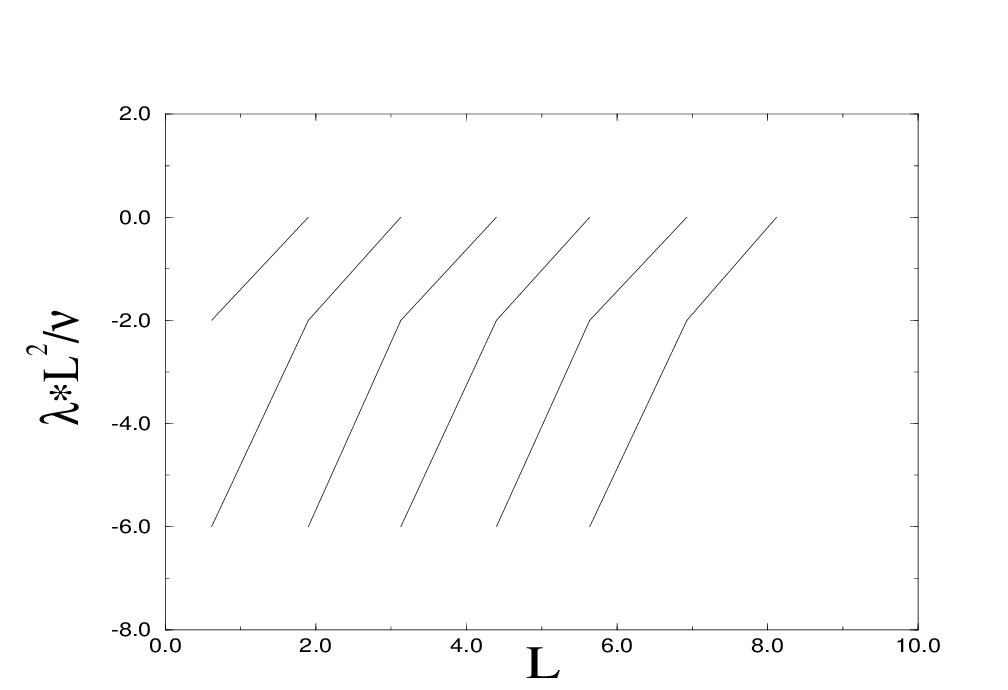}
\caption
{Spectrum of eigenvalues associated with the reaction of the poles
in the giant cusp to the addition of new poles.}
\label{file=Fig.7}
\end{figure}

\end{document}